%% file: ms.tex
\documentclass[sigconf]{acmart}

\usepackage{booktabs} 

\copyrightyear{2019}
\acmYear{2019} 
\setcopyright{iw3c2w3}
\acmPrice{15.00}
\acmDOI{10.1145/1122445.xxxxx}
\acmISBN{978-1-4503-9999-9/18/06}

\usepackage{url}  
\usepackage{graphicx}  
\usepackage{amsmath}
\usepackage{balance}
\usepackage{subcaption}
\usepackage{mwe}
\usepackage{booktabs}
\usepackage{multirow}
\usepackage{tikz}

\usepackage{times}  
\usepackage{helvet}  
\usepackage{courier}  
\usepackage{url}  
\usepackage{graphicx}  
\usepackage{color-edits}
\usepackage{amsmath}
\usepackage{balance}
\usepackage{subcaption}
\usepackage{mwe}
\usepackage{color}
\newcommand{\LL}[1]{\textbf{{\color{blue}(LL: #1)}}}

\usepackage{booktabs} 
\usepackage{subcaption}
\usepackage{tikz}
\usepackage{pgfplots}
\usepackage{makecell}
\usepackage{hyperref}

\usepackage{balance}

\begin{document}

\title{Infringement of Tweets Geo-Location Privacy:\\ an approach based on Graph Convolutional Neural Networks}

\author[L. Luceri]{Luca Luceri}
\thanks{L. Luceri \& D. Andreoletti contributed equally to this work.}
\affiliation{%
  \institution{University of Applied Sciences and Arts of Southern Switzerland, and University of Bern}
  \city{Manno}
  \state{Switzerland}
  \postcode{6928}
}
\email{luca.luceri@supsi.ch}

\author[D. Andreoletti]{Davide Andreoletti}
\affiliation{%
  \institution{University of Applied Sciences and Arts of Southern Switzerland, and Politecnico di Milano}
  \city{Manno}
  \state{Switzerland}
  \postcode{6928}
}
\email{davide.andreoletti@supsi.ch}

\author[S. Giordano]{Silvia Giordano}
\affiliation{%
  \institution{University of Applied Sciences and Arts of Southern Switzerland}
  \city{Manno}
  \state{Switzerland}
  \postcode{6928}
}
\email{silvia.giordano@supsi.ch}

\renewcommand{\shortauthors}{L. Luceri et al.}

\begin{abstract}
The tremendous popularity gained by Online Social Networks (OSNs) raises natural concerns about user privacy in social media platforms.
Though users in OSNs can tune their privacy by deliberately deciding what to share, the interaction with other individuals within the social network can expose, and eventually disclose, sensitive information. 
Among all the sharable personal data, geo-location is particularly interesting. On one hand, users tend to consider their current location as a very sensitive information, avoiding to share it  most of the time.
On the other hand, service providers are interested to extract and utilize geo-tagged data to offer tailored services. In this work, we consider the problem of inferring the current location of a user utilizing only the available information of other social contacts in the OSN. For this purpose, we employ a graph-based deep learning architecture to learn a model between the users' known and unknown geo-location during a considered period of time. 
As a study case, we consider Twitter, where the user generated content (i.e., tweet) can embed user's current location. 
Our experiments validate our approach and further confirm the concern related to data privacy in OSNs.
Results show 
the presence of a critical-mass phenomenon, i.e., if at least 10\% of the users provide their tweets with geo-tags, then the privacy of all the remaining users is seriously put at risk.
In fact, our approach is able to localize almost 50\% of the tweets with an accuracy below 1km relying only on a small percentage of available information.
\end{abstract}

\maketitle

\input{src/conclusion.tex}

\vspace{+.125cm}
\footnotesize{
\textbf{Acknowledgements}. 
The authors gratefully acknowledge support by the Swiss National Science Foundation (SNSF) via the CHIST-ERA project \textit{UPRISE-IoT}.
}
\balance


\bibliographystyle{ACM-Reference-Format}


\end{document}

%% file: src/conclusion.tex
\section{Introduction}

During the past decade, Online Social Networks (OSNs) have gained tremendous popularity worldwide. For example, as of January 2018, Twitter and Facebook count around $330$\footnote{https://www.omnicoreagency.com/twitter-statistics/} and $2170$\footnote{https://wearesocial.com/uk/special-reports/digital-in-2017-global-overview} millions active users, respectively. 
OSNs provide platforms where users can come in contact with each other and share private information about themselves (e.g., interests, age, location, just to name a few). This fact raises natural concerns about privacy issues in OSNs, that users can tune by deliberately deciding what to share. Due to the inherently network-oriented nature of OSNs, however, users interact among each other in several ways, and the information about each user is not completely on its hand. For instance, a generic user can be mentioned in relation to a specific topic by another user, thus revealing a potential interest for a given subject. 
The market value of an OSN highly depends on the amount and quality of data that users share about themselves, as this allows the OSN owner to offer services that are increasingly-tailored toward the particular characteristics of each user. Hence, the OSN operator might be interested to develop tools for extracting as much knowledge as possible from the data of its users. 
An information that users tend to consider particularly sensitive is their location, which OSNs generally allow to associate with the contents that users publish (operation referred to as \textit{geo-tagging}).
 
In this work, we consider the problem of inferring the location associated with users' generated messages published in Twitter (i.e., the \textit{geo-tags} of the \textit{tweets}). Specifically, we aim to infer the locations of tweets without geo-tag from the geo-tagged ones. Notice that we do not base our analysis on the content of the tweets (which is often informative of their locations) but only on signals (i.e., the geo-tags) from the social network users.
The motivation behind this choice is two fold. First, we aim to understand to what extent users' personal information can be estimated from social cues.
Second, we consider the case when user's personal tweets and information are not available (because of privacy settings) and, thus, are not usable inputs for the inference.
Further, we consider the realistic scenario where only a subset of users (not necessarily friends with each other) geo-tag their tweets.

We employ a deep learning architecture to learn a model between the geo-tagged and not-geo-tagged tweets that users generate during a considered period of time. The location is expressed as a pair of latitude and longitude and the inference problem is framed as a regression.  
The employed deep learning architecture has been proposed in \cite{seo2018structured} and it is trained on data structured as sequences of labeled graphs. Specifically, each element of the sequence is a graph representing a snapshot of the OSN in a given period of time, i.e., each node of the network represents a user and its label is the geo-location of the tweet published by the user in the considered period, whereas the edges represent friendship relation between users. The architecture is composed of two main building blocks: a convolutional layer, which learns to represent the relations among the users and a Long-Short-Time-Memory layer, which learns a model of the sequence of graphs. 
Our experiments confirm both the validity of the proposed approach and the serious concern about data privacy in OSN. In fact, one of the main take-aways of our work is the presence of a \textit{critical-mass} phenomenon, i.e., if at least $10\%$ of the users provide their tweets with geo-tags, then the privacy of all the remaining users is seriously put at risk. The results obtained by means of simulations show that with this small percentage of available information our approach is able to localize almost 50\% of the tweets. 

This paper is structured as follows. In Section \ref{rw} we provide some related work about the topics of location inference in Twitter. Section \ref{pd} is devoted to the problem statement. Section \ref{me} describes the proposed system. Results are presented and discussed in Section \ref{ex}, while Section \ref{co} concludes the paper. 

\section{Related Work}
\label{rw}

As one of the most popular OSNs, Twitter has gained a world-wide coverage of users who daily tweet on the social platform. 
Twitter users have the possibility to declare their home addresses and include real-time locations when sharing tweets.
Knowing user location opens the way to several applications.
For this reason, location inference on Twitter has received tremendous interest in the last decade \cite{zheng2018survey}.
Considerable research has focused on this topic - trying to bridge online and offline worlds using location information - developing applications to detect emergency \cite{ao2014estimating,lingad2013location},
monitor public health of citizens \cite{cheng2010you}, recommend places and events \cite{noulas2012mining,yuan2013and}. 

Location in Twitter can be divided in three categories \cite{zheng2018survey}, namely \textit{home location}, \textit{mentioned location}, and \textit{tweet
location}.
Home locations refer to Twitter user's long-term residential address, which can be self-declared (at different levels of granularity) in the user's profile.
Users may also name given locations in the tweet content, thus, these are referred to as mentioned locations.
Finally, tweet location corresponds to the place where a tweet has been generated, which is the geo-tag embedded in the tweet. 
This information is highly valuable for service providers as it allows to have a complete picture of users' mobility, interests, and preferences.
However, only less than 1\% of tweets has explicit geo-tags \cite{graham2014world}. For this reason, tweet location inference remain an open and challenging problem.
As described in \cite{zheng2018survey}, tweet location inference can rely on multiple sources of information: $(i)$ tweet content, $(ii)$ twitter social network among users, and $(iii)$ twitter contextual information, which incorporates the previous two sources along with meta-data related to both tweets and users' profiles. 

In this work, we focus on the tweet location inference, because on one hand users' dynamic geo-location is a desirable information for multiple applications, and on the other hand it represents an open problem for users' privacy in OSNs. In particular, we 
rely only on the source of information provided by the twitter social network. 
Similarly, in \cite{sadilek2012finding}, tweet location inference is performed utilizing only social network information. In particular, the authors model the location sequence of each user with a Dynamic Bayesian Network (DBN), whose features are friends' locations, the time of the day, and the day of the
week. Compared to this approach, our proposed method differs for two main reasons. While in \cite{sadilek2012finding} all the information about users' friends is exploited, we consider the more realistic scenario where only a subset of users (not necessarily friends with each other) provide their location. Moreover, their approach is framed as a classification problem, where nearby locations are merged together in a unique cluster (class), and thus, only known clusters can be inferred. In a different way, we face this problem with a regression model, which infers the pair of location coordinates and can therefore generalize to unseen locations.

\section{Problem Definition}
\label{pd}
\begin{figure*}
        \centering
        \begin{subfigure}[b]{0.32\textwidth}
               \includegraphics[width=\textwidth]{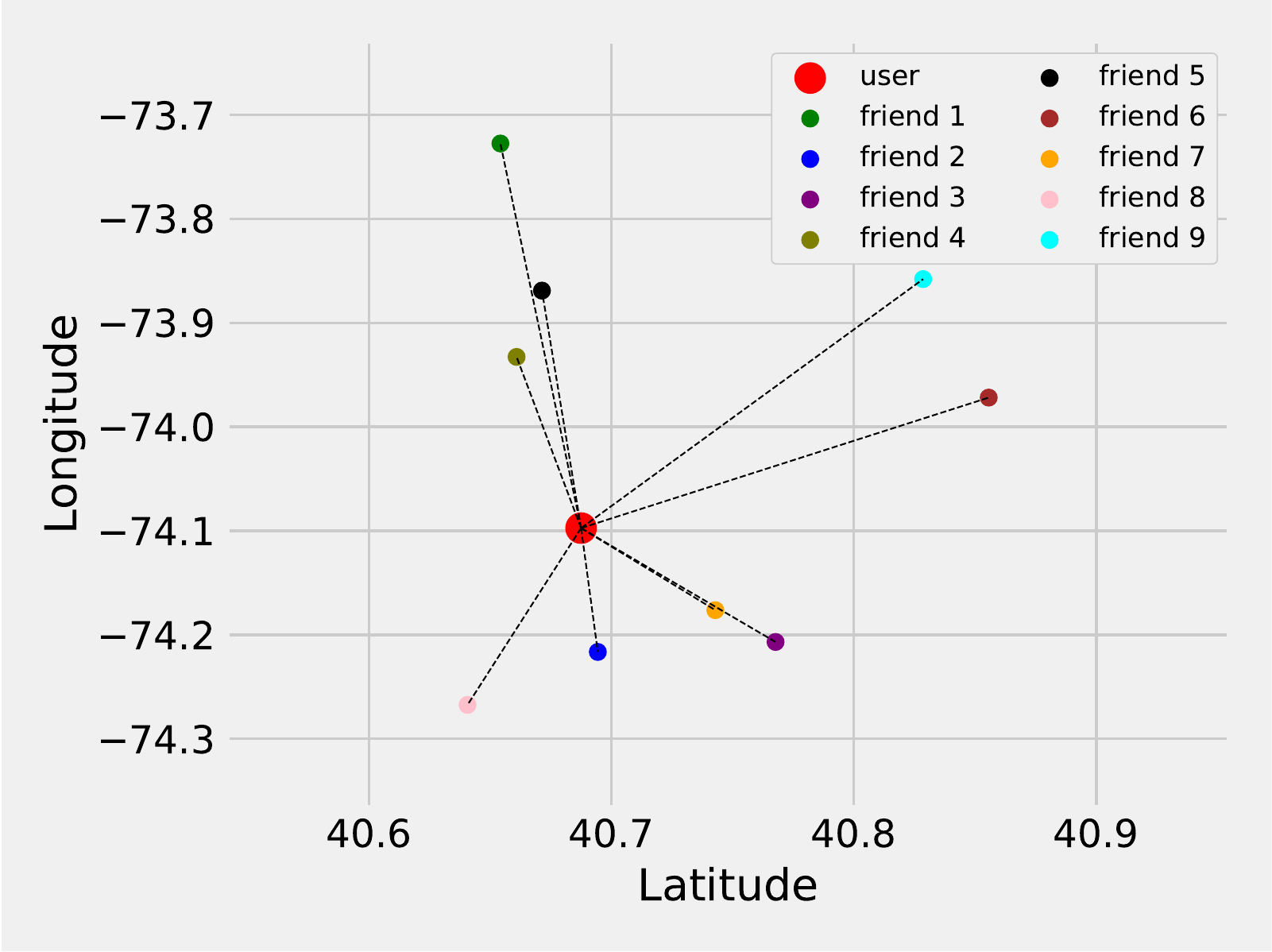}
                \caption{time slot 120}
                \label{fig1a}
        \end{subfigure}\hfill%
        \begin{subfigure}[b]{0.32\textwidth}
               \includegraphics[width=\textwidth]{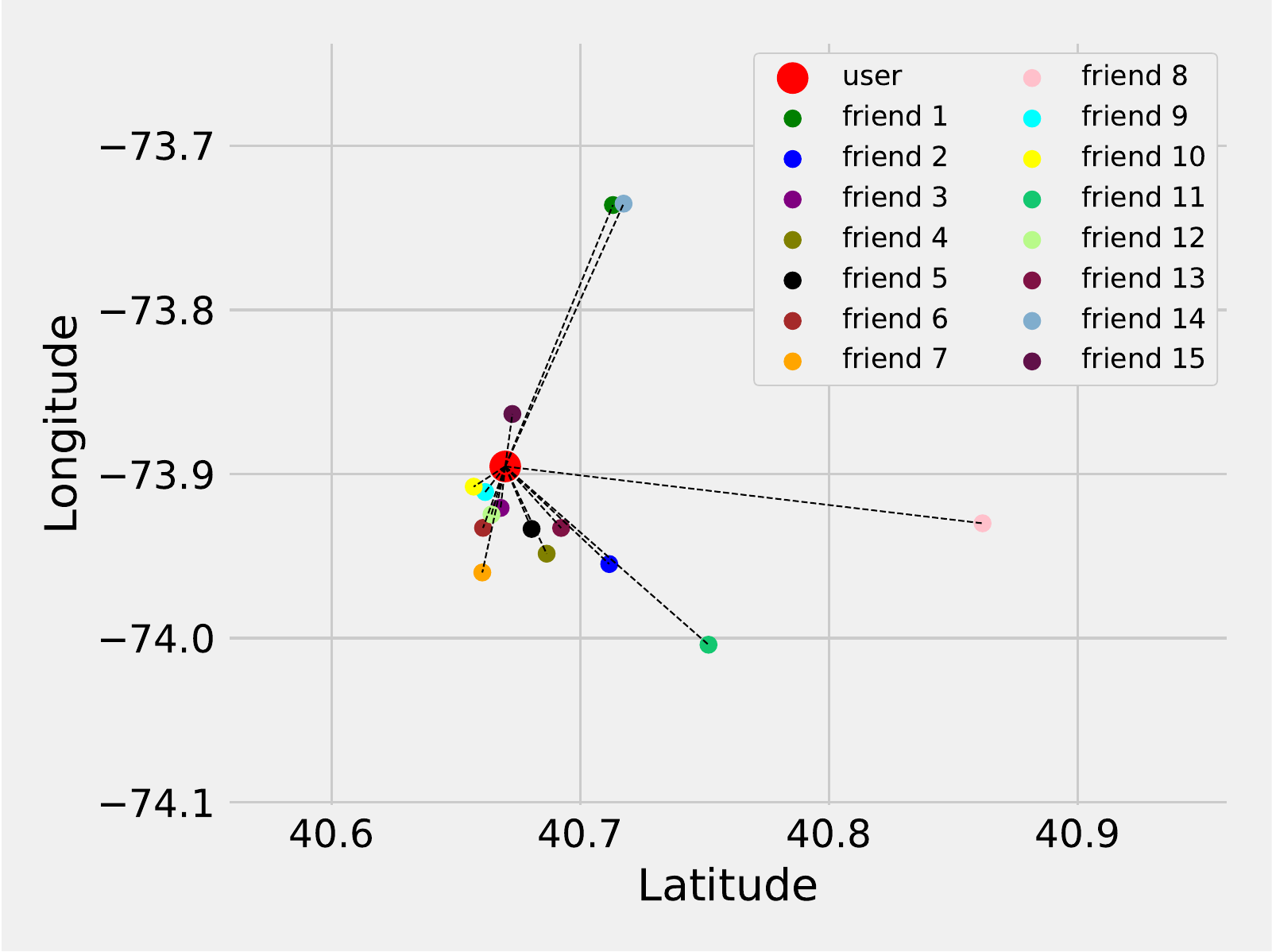}
                \caption{time slot 305}
                \label{fig1b}
        \end{subfigure}\hfill%
        \begin{subfigure}[b]{0.32\textwidth}
                \includegraphics[width=\textwidth]{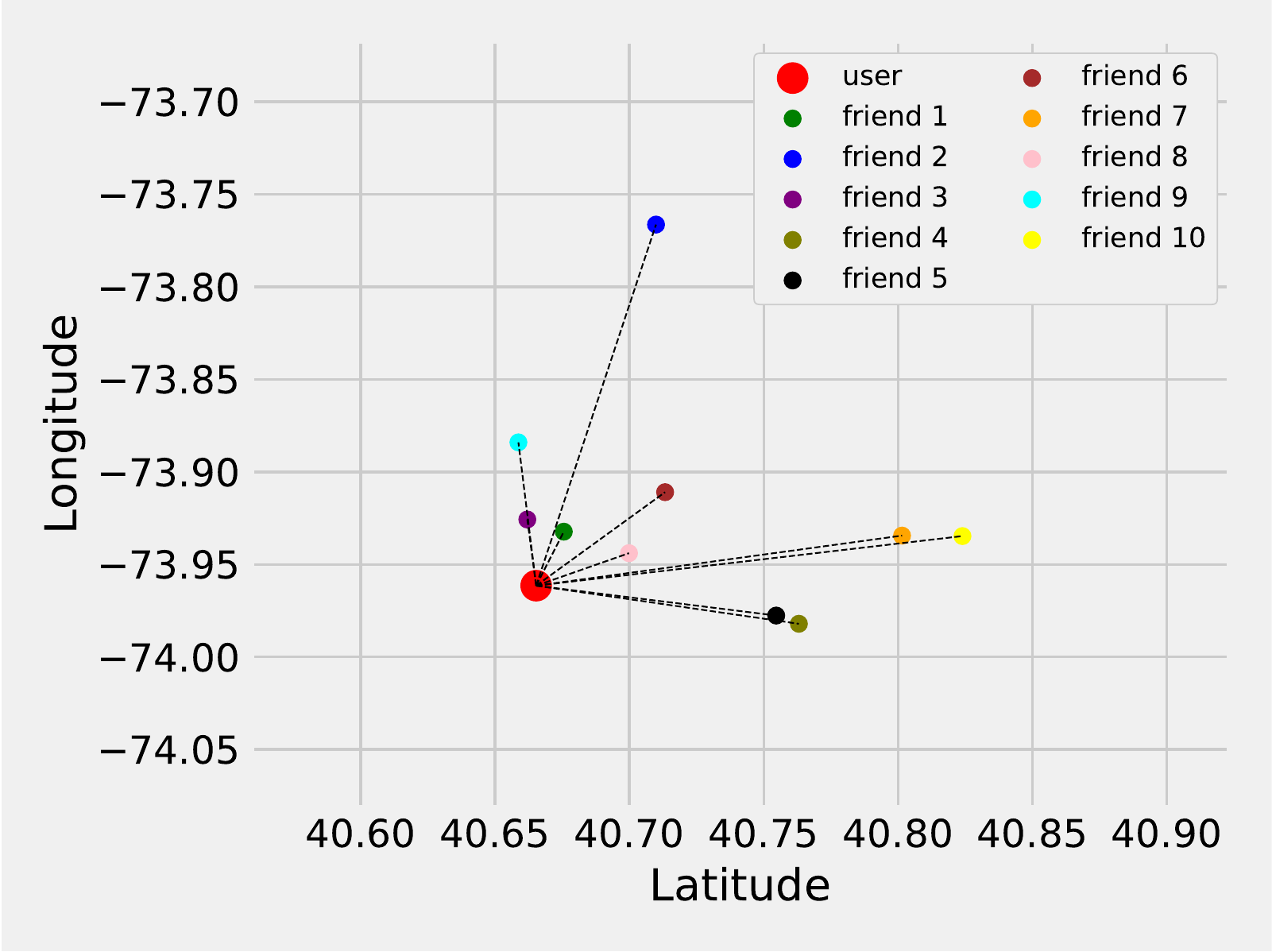}
                \caption{time slot 430}
                \label{fig1c}
        \end{subfigure}      
        \caption{GEO-SN of a given user at different time slots.}
        \label{prob_statement}
\end{figure*}

Let $G = (V,E)$ be a directed graph representing Twitter social network, where $V=\{u_1,u_2,\dots,u_N\}$ is the set of users, and E is the set of edges that link them.
Connections among users in Twitter are based on the \textit{followee/follower} paradigm. A generic user $u_{i}$ can follow another user $u_{j}$ without being necessarily followed back. For this reason, we consider $u_j$ and $u_i$ to be \textit{friends} iff $(u_j,u_i) \in E$ and $(u_i,u_j) \in E$. 

We define $\mathcal{T} = \{\tau_{1},\tau_{2}, ..., \tau_{M}\}$ as the set of available tweets posted on Twitter within a considered period of time. 
In Twitter, users can provide information about their current location by explicitly geo-tagging their tweet in the OSN platform. 
However, 
users do not disclose their location in most of their tweets.
Though hiding this information may preserve user's privacy, it has been shown that location can be inferred combining multiple sources of information, e.g., social network, tweet text, and mentions.

In this paper, we investigate how to discover the hidden location information in users' tweet exploiting only social cues.
We propose to infer user's geo-location by leveraging on social contacts available information in a given time slot. 
Time is discretized into slots of duration $\Delta t$ in order to have as much information as possible from several users in a limited amount of time.
As an example, Figure \ref{prob_statement} shows the locations of a user and her friends at different time slots. 
User location is defined by the pair of coordinates (latitude, longitude). 
Connections represent distances between the user and her friends in the given time slot.
We refer to this dynamic topology as \textit{GEO-Social Network} (GEO-SN).
GEO-SN is a geo-spatial network as the position of the nodes indicates a precise geo-location. 
Moreover, GEO-SN is dynamic as $(i)$ the locations and the distances between the user and her friends vary with time and $(ii)$ the number of nodes changes according to the number of friends providing their location within the time slot, as can be appreciated from Figure \ref{prob_statement}.
For the sake of simplicity, in Fig. \ref{prob_statement}, we show only the social network of a single user, but our approach deals with the whole social network topology.


Formalizing the problem, our objective is to determine $\mathbf{l}^{t_m}_{u_{i}}$, the geo-location of user $u_i$ at time slot $t_{m}$,  $\forall i \in V$ and $\forall m \in \{1,2, ..., M\}$, exploiting the location (if known) of other users in the OSN, i.e., $\{ \mathbf{l}^{t_m}_{u_{j}},  \mathbf{l}^{t_m}_{u_{k}}, \dots,  \mathbf{l}^{t_m}_{u_{n}} \}$.
Overall, we aim to find a function $f$ that models user movements within the GEO-SN.
This function should be able to map each user's location with others' location by learning spatial and temporal dependencies among them.
Therefore, we define $f(\mathbf{x})$ as
\begin{equation}
  f(\mathbf{x})=f(\{ \mathbf{l}^{t_m}_{u_{j}},  \mathbf{l}^{t_m}_{u_{k}}, \dots,  \mathbf{l}^{t_m}_{u_{n}} \})=\hat{\mathbf{l}}^{t_m}_{u_{i}},
\end{equation}
where $\hat{\mathbf{l}}^{t_m}_{u_{i}}$ is the predicted location of $u_i$ at time slot $t_m$.

\section{Methodology}
\label{me}

In this Section, we present the proposed methodology based on a deep learning architecture, which is discussed in turn.
Our purpose is to model the spatial and temporal dependencies that users within an OSN have on each other in relation to the information about their tweets' geo-tags. Specifically, our aim is to use this model to infer the most probable location of a target user tweet based only on other users' available geo-tags.  
We consider the temporal dependencies among tweets' location and the social relations among users as essential factors of this model, that we obtain by following a deep-learning approach. The nodes of a graph are characterized by i) a set of attributes, ii) a set of labels and iii) a relation among each other. 
 
While i) and ii) are common characteristics of data used with machine learning, the question on how to learn a model also based on the relations among the nodes of a graph has been receiving a significant focus recently. A viable approach to consider the relations among users may consist in processing the graph in such a way that each node's attributes are enriched with its neighbors attributes (operation often referred to as \textit{embedding}). This approach, however, models only the relations among groups of friends within an OSN and, thus, does not fit our scenario, where only a subset of users (not necessarily friends with each other) provide their location.
Therefore, we are interested in modeling complex relations that go beyond the friendship among users to learn spatio-temporal dependencies among them.

An existing deep learning architecture that suites our needs has been proposed in \cite{seo2018structured}. This architecture is referred to as GCNN-LSTM and it is based on two main building blocks, namely a convolutional layer and a recurrent layer. The former is aimed to extract relevant social patterns, whereas the latter learns a temporal model of the sequences of geo-tags. The deep learning architecture is obtained by stacking the recurrent layer on top of the convolutional one. 
The recurrent module implemented using the LSTM \cite{hochreiter1997long} captures long-term dependencies among the elements of the sequence (i.e., the graphs at different time slots). However, this module may fall short in modeling short-term relations within the sequence. To capture both short and long-term relations we also make use of a CNN-1D, which proved successful in modeling sequences (e.g., sentences \cite{kalchbrenner2014convolutional}).
The resulting architecture is composed of a CNN-1D, a set of GCNN-LSTMs and a fully-connected layer stacked on each other. The CNN-1D is characterized by a number of filters $N_{CNN}$ of size $w_{CNN}$. The GCNN-LSTM and the fully-connected layers are characterized by their number of neurons, i.e., $N_{G}$ and $N_{F}$, respectively.   

\section{Experiments}
\label{ex}
\subsection{Data}

\begin{figure}[!t]
\centering
\includegraphics[width=3in]{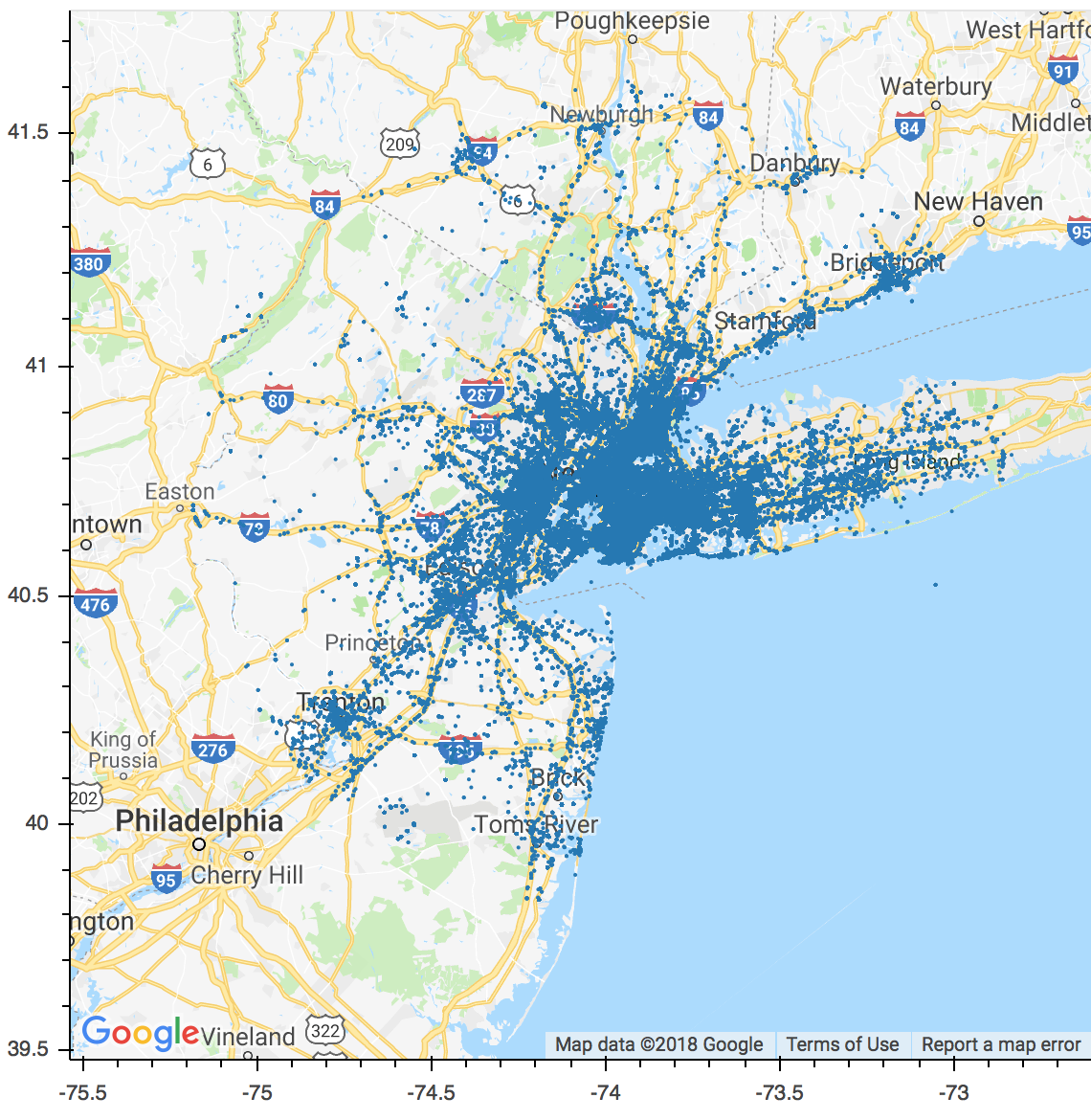} 
\caption{NYC Tweets in 2010}
\label{map_nyc2010}
\end{figure}

To validate and evaluate our approach we make use of the Twitter dataset collected in  \cite{sadilek2012finding}, which gathered tweets within 100 kilometers of New York city center for 31 days.
Figure \ref{map_nyc2010} shows the spatial distribution of the tweets over all the collection period.
Social network connections have been collected, along with geo-tagged tweets, and utilized to reveal friendship relations among users.

In Table \ref{data_stats}, we summarize the statistics about the data and network properties related to the social graph.
The \textit{average degree} is the average number of friends over all the users in the OSN, while the \textit{diameter} is the longest of the shortest paths in the social network. The \textit{clustering coefficient} is the average of the clustering coefficients over all the users, where the latter is the ratio between the number of links connecting user's friends to each other to the possible number of possible connections. Finally, the \textit{density} is the ratio between the number of edges connecting the users and the number of possible edges in a network with $N$ users. 


\begin{table}
\centering
\caption{Basic statistics of the Twitter dataset used in this study.} 
\begin{tabular}{ccc}
\hline
& New York City Dataset    \\
\hline
Unique users      &6082  \\  
Friendship relationships      &31874 \\  
Average Degree      &10.22 \\  
Diameter     &19 \\  
Clustering Coefficient   &0.15 \\  
Density      &0.001 \\  
Tweets   &2173681  \\  
Locations      &47808   \\  
\hline
\end{tabular}
\label{data_stats}
\end{table}

\subsection{Simulation Settings}
\label{sim_set}
Our objective is to evaluate the ability to violate users' privacy at a given time slot $t$ given the information on the past $N_{TS}$ graphs. Each graph represents a snapshot of the OSN in a given time slot of duration $\Delta t = 3$ hours\footnote{We set this parameter based on the average time between two consecutive tweets in the dataset.}.
Specifically, the input of our deep learning architecture is the sequence of the graphs at time steps $\left[t -N_{TS} +1, ..., t\right]$, where at each time step we consider a different set of users who are providing the location to their tweets. The output is the inference of the location of those users who have not provided their location at time $t$.  
We process the data in order to make them suitable to feed the deep learning model. This processing results in input shaped as 3D tensors of dimensions $(N_{TS}, N, F)$, where $N_{TS}$ is the number of time-steps (i.e., how far we look in the past of our sequence of graphs), the total number of users $N$ and $F$ being the number of features of each data point. For those users who do not publish a tweet at a given time slot, we consider their last available geo-tags as the features of the corresponding nodes in that time slot. Hence, the number of features is set to $F = 3$ to account for i) latitude, ii) longitude and iii) distance (in number of time slots) between the considered time slot and the last time slot where the user has provided a valid geo-tag.  
On the other hand, the output are matrices of dimension $(N,M)$, where $M = 2$ is the number of outputs (i.e., to account for inferred latitude and longitude). We then normalize the input and output tensors between $0$ and $1$. 

Our primary objective is to assess the ability to violate users' privacy as a function of the percentage of tweets that are geo-tagged. To this aim, we introduce the parameter $p$ as the probability that a tweet is geo-tagged. We expect that low values of $p$ will make the inference less effective, since the deep learning architecture is trained on less data and the learned model is then prone to over-fitting issues. 
Conversely, high values of $p$ will result in more information available for the OSN provider and, consequently, will lead to an increased ability to violate users' privacy. 
We divide our data into three non-overlapping sets, namely \textit{training}, \textit{validation}, and \textit{test} sets. Specifically, at each time slot we assign each node of the graph to one of these sets. A node belongs to the training, validation and test set with probabilities $p$, $0.5\cdot(1-p)$ and $0.5\cdot(1-p)$, respectively. Notice that we perform the training, validation and test considering only the users for which, at each time slot, the next location is known. The others are simply discarded. 

We employ a deep learning architecture formed by stacking together a CNN-1D with four filters ($N_{CNN}=4$) of size $w_{CNN}=4$, three GCNN-LSTMs with $N_{G}$=$\{20,10,30\}$ neurons, respectively, and a fully-connected layer of $N_{F}=2$ neurons (which encode the inferred latitude and longitude). 

\subsection{Results}
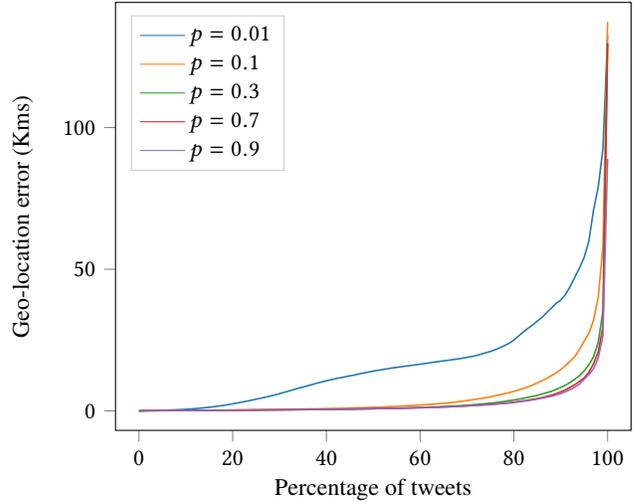
\begin{figure}{
\label{fig:percentile}
\centering 

\begin{tikzpicture}

\definecolor{color0}{rgb}{0.12156862745098,0.466666666666667,0.705882352941177}
\definecolor{color1}{rgb}{1,0.498039215686275,0.0549019607843137}
\definecolor{color2}{rgb}{0.172549019607843,0.627450980392157,0.172549019607843}
\definecolor{color3}{rgb}{0.83921568627451,0.152941176470588,0.156862745098039}
\definecolor{color4}{rgb}{0.580392156862745,0.403921568627451,0.741176470588235}

\begin{axis}[
legend cell align={left},
legend entries={{$p = 0.01$},{$p = 0.1$},{$p = 0.3$},{$p = 0.7$},{$p = 0.9$}},
legend style={at={(0.03,0.97)}, anchor=north west, draw=white!80.0!black},
tick align=outside,
tick pos=left,
x grid style={white!69.01960784313725!black},
xlabel={Percentage of tweets},
xmin=-5, xmax=105,
y grid style={white!69.01960784313725!black},
ylabel={Geo-location error (Kms)},
ymin=-6.86482051750062, ymax=144.179941791791
]
\addlegendimage{no markers, color0}
\addlegendimage{no markers, color1}
\addlegendimage{no markers, color2}
\addlegendimage{no markers, color3}
\addlegendimage{no markers, color4}
\addplot [semithick, color0]
table [row sep=\\]{%
0	0.000850496558069495 \\
1	0.0626999015185792 \\
2	0.105450992590022 \\
3	0.144158145138653 \\
4	0.189325020684946 \\
5	0.233815580004995 \\
6	0.283171937782493 \\
7	0.342407585965223 \\
8	0.410480336574793 \\
9	0.481808794974339 \\
10	0.574670949158113 \\
11	0.677812003645663 \\
12	0.800163914089727 \\
13	0.946348666475448 \\
14	1.10925050518283 \\
15	1.28346957910584 \\
16	1.49845425418851 \\
17	1.71416020154096 \\
18	1.94983962107673 \\
19	2.19838753429506 \\
20	2.49447336983858 \\
21	2.79860671974724 \\
22	3.1012375543115 \\
23	3.4333334213962 \\
24	3.76579559305467 \\
25	4.1324658605245 \\
26	4.50006334225291 \\
27	4.85023226856686 \\
28	5.24687424029162 \\
29	5.63263627903302 \\
30	6.09926177825714 \\
31	6.56487283058482 \\
32	7.01491841566125 \\
33	7.49997889063673 \\
34	8.0069961581778 \\
35	8.41180531573355 \\
36	8.8820804588658 \\
37	9.30464935306327 \\
38	9.78276141043869 \\
39	10.2049864957684 \\
40	10.643692937119 \\
41	11.0048988742528 \\
42	11.3767049419466 \\
43	11.7149033847316 \\
44	12.0429245283435 \\
45	12.3896377840966 \\
46	12.6782036146869 \\
47	13.0613169103933 \\
48	13.3999308698387 \\
49	13.7118281055508 \\
50	14.0574590169479 \\
51	14.3116111492934 \\
52	14.6006443662685 \\
53	14.8821663045532 \\
54	15.1398490731354 \\
55	15.3660896047318 \\
56	15.5996791060101 \\
57	15.8277694411866 \\
58	16.0315281952463 \\
59	16.2565902440883 \\
60	16.4924004006908 \\
61	16.7237476955929 \\
62	16.956630844182 \\
63	17.2167259575957 \\
64	17.438177027164 \\
65	17.6626679709814 \\
66	17.870727576207 \\
67	18.0988465984083 \\
68	18.3272577870686 \\
69	18.5993338850579 \\
70	18.8801983645298 \\
71	19.1978903388322 \\
72	19.5585661132931 \\
73	20.0550429969691 \\
74	20.5198608887727 \\
75	20.9854023063624 \\
76	21.581340356838 \\
77	22.2894925349958 \\
78	23.0296357300113 \\
79	23.873319900659 \\
80	24.9769929924516 \\
81	26.4384862140961 \\
82	27.7889030418615 \\
83	29.1552488824196 \\
84	30.2914443530615 \\
85	31.7239440914696 \\
86	32.9540112766612 \\
87	34.700076293876 \\
88	36.2889344413985 \\
89	37.9484388021693 \\
90	39.1322772159256 \\
91	41.0492106581752 \\
92	43.8915046115686 \\
93	47.1797304548199 \\
94	50.4696182613916 \\
95	54.1812214127954 \\
96	59.8438633841725 \\
97	70.924773208319 \\
98	79.0031839922428 \\
99	92.4663085103486 \\
100	132.625766973304 \\
};
\addplot [semithick, color1]
table [row sep=\\]{%
0	0.00176336415187122 \\
1	0.0466332669639476 \\
2	0.0672921658784802 \\
3	0.0839826205447238 \\
4	0.0999431473868296 \\
5	0.114224837892678 \\
6	0.127619593247135 \\
7	0.141671983052894 \\
8	0.155315407533968 \\
9	0.167881389234864 \\
10	0.182184521588196 \\
11	0.195584368048343 \\
12	0.210495905247251 \\
13	0.224245497269764 \\
14	0.238988617671626 \\
15	0.252842911662427 \\
16	0.268026390509343 \\
17	0.284346898595307 \\
18	0.299932758195696 \\
19	0.315643000894221 \\
20	0.332350463410266 \\
21	0.350133944895575 \\
22	0.367296687438233 \\
23	0.383733302208889 \\
24	0.402631058117167 \\
25	0.420913587386746 \\
26	0.441201404413458 \\
27	0.461457371607591 \\
28	0.483873109529638 \\
29	0.504854327705791 \\
30	0.526342906449404 \\
31	0.548073090577101 \\
32	0.573388425323946 \\
33	0.600229972958979 \\
34	0.626952754971318 \\
35	0.654284739897358 \\
36	0.683381321977565 \\
37	0.715179853494969 \\
38	0.743279851914358 \\
39	0.777226691513721 \\
40	0.811740541597859 \\
41	0.847627775612144 \\
42	0.884403993265156 \\
43	0.92540688050194 \\
44	0.966801191847993 \\
45	1.01189769223572 \\
46	1.05737348521022 \\
47	1.10624414868599 \\
48	1.15755067380845 \\
49	1.21354864563994 \\
50	1.26841109478398 \\
51	1.32511474436942 \\
52	1.38779996843076 \\
53	1.46114856331822 \\
54	1.53751704838672 \\
55	1.61008188860645 \\
56	1.69218729474998 \\
57	1.78052741084892 \\
58	1.88102635146621 \\
59	1.97590093234113 \\
60	2.07157489819801 \\
61	2.18587793514719 \\
62	2.30955878396917 \\
63	2.42300236830314 \\
64	2.56168320268448 \\
65	2.6959321296079 \\
66	2.85448146386319 \\
67	3.02973252306104 \\
68	3.21674185510051 \\
69	3.42452772024234 \\
70	3.65321926852271 \\
71	3.8879938144481 \\
72	4.13941227625525 \\
73	4.39402110671232 \\
74	4.67160222106061 \\
75	4.98494041936292 \\
76	5.33581230777364 \\
77	5.70357656103933 \\
78	6.0670095828741 \\
79	6.46632680524892 \\
80	6.91312564792687 \\
81	7.37344378429745 \\
82	7.89865923964394 \\
83	8.45834502995514 \\
84	9.12262904838628 \\
85	9.88687174205472 \\
86	10.6322590012311 \\
87	11.4764024848482 \\
88	12.4135158149433 \\
89	13.4135165404881 \\
90	14.5749271285048 \\
91	16.0173556656093 \\
92	17.5195069909897 \\
93	19.1819616894252 \\
94	21.7705113213135 \\
95	24.5861757373282 \\
96	27.3947010133432 \\
97	32.0264438050618 \\
98	40.1150156591347 \\
99	57.9733679201066 \\
100	137.314270777732 \\
};
\addplot [semithick, color2]
table [row sep=\\]{%
0	0.00401643149830618 \\
1	0.0394932480139353 \\
2	0.0560685100850663 \\
3	0.0700538986454194 \\
4	0.0817449724711345 \\
5	0.0913680628185631 \\
6	0.100687208005054 \\
7	0.110943404411289 \\
8	0.120876531233314 \\
9	0.130100254453276 \\
10	0.140391367719066 \\
11	0.149059804936389 \\
12	0.158395543400389 \\
13	0.168511251676598 \\
14	0.177438722318271 \\
15	0.186715003886265 \\
16	0.196857596408781 \\
17	0.206950173155968 \\
18	0.217171918887366 \\
19	0.22853828338566 \\
20	0.239364192590256 \\
21	0.251022257419227 \\
22	0.261890603770665 \\
23	0.273546542989515 \\
24	0.285003740562359 \\
25	0.297260730551554 \\
26	0.309198015755265 \\
27	0.32171033516562 \\
28	0.334381928808336 \\
29	0.348383129413481 \\
30	0.363772011164784 \\
31	0.37931768150687 \\
32	0.394715464086132 \\
33	0.409794325318983 \\
34	0.425826785107594 \\
35	0.442661006463901 \\
36	0.460207862494528 \\
37	0.477983331100469 \\
38	0.496369192175353 \\
39	0.514660160556423 \\
40	0.534400918170183 \\
41	0.557072321836314 \\
42	0.578486424362496 \\
43	0.5990934229656 \\
44	0.621138427232764 \\
45	0.643974423743866 \\
46	0.668726191151841 \\
47	0.695265726195388 \\
48	0.723984642911485 \\
49	0.752535934211337 \\
50	0.783240904318704 \\
51	0.814236407098857 \\
52	0.85213317554739 \\
53	0.890908755786837 \\
54	0.92697982752134 \\
55	0.964909610305607 \\
56	1.01006624095352 \\
57	1.05517018002785 \\
58	1.10581002295479 \\
59	1.15936240519273 \\
60	1.21169321066836 \\
61	1.27125322251241 \\
62	1.33059812392368 \\
63	1.40289962501614 \\
64	1.47800588732068 \\
65	1.55153043166681 \\
66	1.62735276183672 \\
67	1.71271392016224 \\
68	1.79417369628249 \\
69	1.90053723541497 \\
70	2.01491030750656 \\
71	2.13339912983697 \\
72	2.25792921112584 \\
73	2.39828048866771 \\
74	2.56600687205822 \\
75	2.73730237011965 \\
76	2.91655257910628 \\
77	3.12372530452462 \\
78	3.35974031133763 \\
79	3.60696817362762 \\
80	3.84812524204972 \\
81	4.13233131063297 \\
82	4.42345570326344 \\
83	4.77128342576821 \\
84	5.11698444200407 \\
85	5.47696439133242 \\
86	5.95356950698925 \\
87	6.42359183792433 \\
88	6.92047383670468 \\
89	7.56180658256616 \\
90	8.29624109666794 \\
91	9.05414844527247 \\
92	10.0411806685628 \\
93	11.1524709692043 \\
94	12.4490999514816 \\
95	14.0965471696338 \\
96	16.2957052907852 \\
97	19.0665306330491 \\
98	24.2650414953865 \\
99	35.0777109022963 \\
100	129.87802687933 \\
};
\addplot [semithick, color3]
table [row sep=\\]{%
0	0.00193759603004313 \\
1	0.0388917785898547 \\
2	0.0562023471285048 \\
3	0.0701928350305017 \\
4	0.0831829823591985 \\
5	0.0961861357004997 \\
6	0.108179971643105 \\
7	0.118381078443624 \\
8	0.128040551967899 \\
9	0.137912256135109 \\
10	0.148048591073777 \\
11	0.158218723819117 \\
12	0.167737936637293 \\
13	0.17740918250177 \\
14	0.18822568409638 \\
15	0.198820519547656 \\
16	0.209585340961115 \\
17	0.219302584883652 \\
18	0.228896210917525 \\
19	0.238183312957222 \\
20	0.247860295592289 \\
21	0.257800513872119 \\
22	0.268641352314491 \\
23	0.280061161474976 \\
24	0.291024611774657 \\
25	0.302132992694175 \\
26	0.313367039295743 \\
27	0.326057653627681 \\
28	0.339076727724654 \\
29	0.350093397359794 \\
30	0.362457314729072 \\
31	0.376079049005318 \\
32	0.391522352707084 \\
33	0.407233713061006 \\
34	0.42234348080869 \\
35	0.438499613161961 \\
36	0.455060305978266 \\
37	0.470962527587973 \\
38	0.490163472775936 \\
39	0.507698735559972 \\
40	0.527023500118863 \\
41	0.546977722184034 \\
42	0.566536965682993 \\
43	0.588103388550738 \\
44	0.609565508704987 \\
45	0.634923801429694 \\
46	0.660522275746606 \\
47	0.685860261731692 \\
48	0.71062389069686 \\
49	0.734830617314959 \\
50	0.767035215304371 \\
51	0.800060280791832 \\
52	0.831592297742361 \\
53	0.862965792607114 \\
54	0.896432355026539 \\
55	0.932695784836577 \\
56	0.96912881839297 \\
57	1.00616653254411 \\
58	1.04414422464151 \\
59	1.08285370934834 \\
60	1.12851907719278 \\
61	1.17337894474332 \\
62	1.21754677950434 \\
63	1.26820790592654 \\
64	1.33074690221652 \\
65	1.39287764085124 \\
66	1.4589892139601 \\
67	1.54156129461806 \\
68	1.61752590515537 \\
69	1.69569477768163 \\
70	1.78956127635407 \\
71	1.87753356120387 \\
72	1.98556796447199 \\
73	2.09807604261721 \\
74	2.22396367222252 \\
75	2.34775176859033 \\
76	2.45918272493674 \\
77	2.59092555011685 \\
78	2.7363530149374 \\
79	2.90046228588674 \\
80	3.11118321738048 \\
81	3.33229113779277 \\
82	3.55619628833545 \\
83	3.81129918089479 \\
84	4.06558824406926 \\
85	4.37002216431657 \\
86	4.70777562343379 \\
87	5.11854235393247 \\
88	5.54328826728747 \\
89	6.07495544506444 \\
90	6.72118170345013 \\
91	7.43854102343453 \\
92	8.33252531077575 \\
93	9.25004055269443 \\
94	10.4170963140536 \\
95	11.6794067437987 \\
96	13.6256799769617 \\
97	16.7006587981852 \\
98	20.8020970130623 \\
99	29.7077674755862 \\
100	129.664419972285 \\
};
\addplot [semithick, color4]
table [row sep=\\]{%
0	0.00185627763450269 \\
1	0.0510428845877365 \\
2	0.0710873390191381 \\
3	0.0874148943779246 \\
4	0.0979161581680278 \\
5	0.110134716416771 \\
6	0.118641028137991 \\
7	0.129302053927618 \\
8	0.139665868835432 \\
9	0.15173829971774 \\
10	0.16116143409049 \\
11	0.171254778565886 \\
12	0.179669294055072 \\
13	0.190708956722328 \\
14	0.20044688611634 \\
15	0.209811405212364 \\
16	0.21895323084722 \\
17	0.22930576382373 \\
18	0.240188307509112 \\
19	0.25062343691488 \\
20	0.261137254594752 \\
21	0.271137287745438 \\
22	0.278510321354323 \\
23	0.289455142698011 \\
24	0.301674839796204 \\
25	0.313337775976691 \\
26	0.324856987159544 \\
27	0.340657611048789 \\
28	0.353985137656765 \\
29	0.369074861105079 \\
30	0.381784995251913 \\
31	0.396743841910133 \\
32	0.410096626328024 \\
33	0.422909491888558 \\
34	0.435600843256366 \\
35	0.4523763671805 \\
36	0.470783853785687 \\
37	0.484986428127149 \\
38	0.503853162688969 \\
39	0.521344541847102 \\
40	0.541791745208932 \\
41	0.56160598801162 \\
42	0.578117225796832 \\
43	0.600755693836523 \\
44	0.629823441114243 \\
45	0.647967608721029 \\
46	0.671603654423297 \\
47	0.697896741370662 \\
48	0.72043192880447 \\
49	0.740154903428007 \\
50	0.761888903799238 \\
51	0.79385363580011 \\
52	0.819397895138384 \\
53	0.848113705296224 \\
54	0.879023839960918 \\
55	0.909845558249256 \\
56	0.949258282226656 \\
57	0.977915893648928 \\
58	1.01342101892186 \\
59	1.04712099366145 \\
60	1.0862760301824 \\
61	1.13028941693206 \\
62	1.17455310995707 \\
63	1.22564615479015 \\
64	1.2752708711329 \\
65	1.33224585478014 \\
66	1.40524405760861 \\
67	1.47577725126024 \\
68	1.54635698570183 \\
69	1.6191213549267 \\
70	1.73167086583591 \\
71	1.83381629178352 \\
72	1.90740807573965 \\
73	2.00351137171857 \\
74	2.10425875896549 \\
75	2.19567212316025 \\
76	2.3475798452122 \\
77	2.48712962962389 \\
78	2.64001858849265 \\
79	2.7782211959069 \\
80	2.96621455671106 \\
81	3.16606348287882 \\
82	3.40624935746413 \\
83	3.61641656896758 \\
84	3.85414606268215 \\
85	4.17116438785469 \\
86	4.50845824100604 \\
87	4.84277076180141 \\
88	5.15238705644067 \\
89	5.61185112592489 \\
90	6.09318036622776 \\
91	6.77577446728162 \\
92	7.5003706079312 \\
93	8.399737115182 \\
94	9.46194128774412 \\
95	11.1755096139568 \\
96	12.8365777818712 \\
97	14.8630041272574 \\
98	18.691783612645 \\
99	26.5707189700546 \\
100	89.0288908123321 \\
};
\end{axis}

\end{tikzpicture}

\caption{Percentile of the geo-localization error (in Kms) computed over the test with varying probability $p$ that a tweet is geo-tagged}\label{fig:percentile}
}\end{figure}


\begin{figure*}
         \centering
         \begin{subfigure}[b]{0.31\textwidth}
                \includegraphics[width=\textwidth]{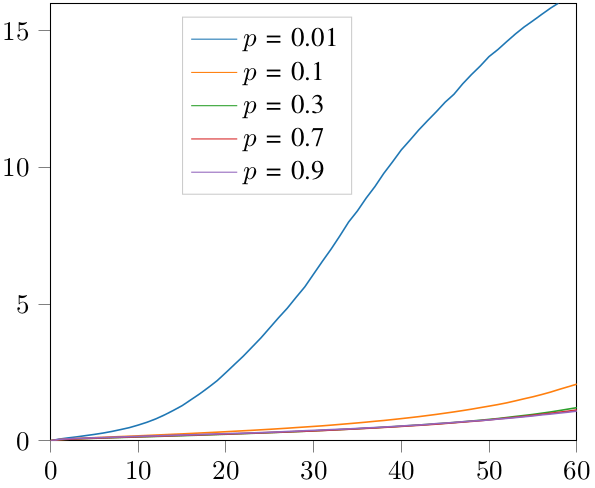}
                 \caption{Highly-predictable tweets}
                 \label{fig:high}
         \end{subfigure}\hfill%
         \begin{subfigure}[b]{0.31\textwidth}
                \includegraphics[width=\textwidth]{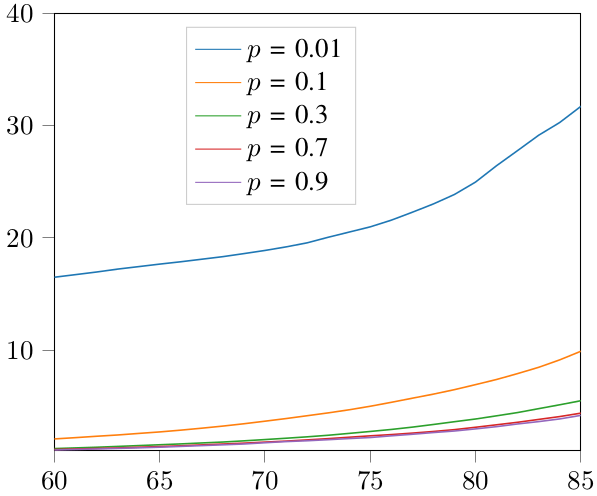}
                 \caption{Average-predictable tweets}
                 \label{fig:average}
         \end{subfigure}\hfill%
         \begin{subfigure}[b]{0.31\textwidth}
                 \includegraphics[width=\textwidth]{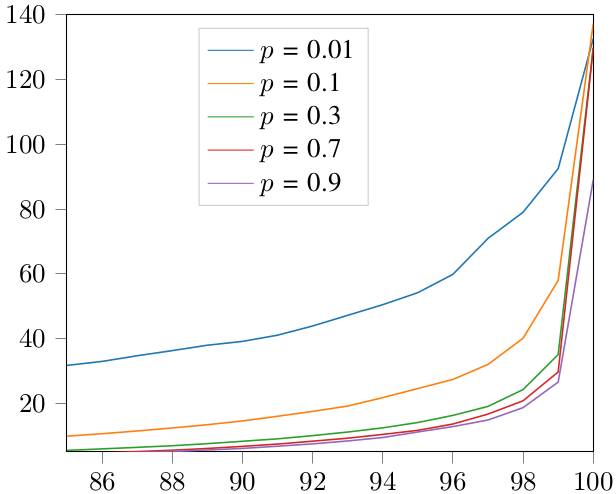}
                 \caption{Poorly-predictable tweets}
                 \label{fig:poorly}
         \end{subfigure} 
         \caption{Three main regions of the percentile of the geo-localization error (in Kms) with varying probability $p$ that a tweet is geo-tagged, computed on the test set}
         \label{fig:percentile_zoom}
\end{figure*}

\begin{table}[t]
\caption{Mean values and percentage of tweets geo-localized with an error $< 1$km, for different values of geo-tag probability $p$}
\centering 
\begin{tabular}{|c|c|c|}
\hline
         & \thead{\textbf{Mean Error} \\ \textbf{(in Kms)}} & \thead{\textbf{\% of tweets geo-localized} \\ \textbf{with an error < 1 km}} \\ \hline
$p=0.01$  & $17.07$                                       & $13.36\%$                                                                         \\ \hline
$p=0.1$ & $5.28$                                        & $45.20\%$                                                                         \\ \hline
$p=0.3$ & $3.13$                                        & $56.53\%$                                                                          \\ \hline
$p=0.7$ & $2.68$                                        & $57.17\%$                                                                         \\ \hline
$p=0.9$ & $2.57$                                        & $57.26\%$                                                                         \\ \hline
\end{tabular}
\label{tab:mean1km}
\end{table}

In this Section we present an overview of the results that we have obtained. We recall that we evaluate the proposed model on graph-structured data, i.e., the sequence of the graphs representing the OSN at each time slot. At each time slot, graphs' nodes are characterized by a unique attribute and a unique label. As explained in \ref{sim_set}, the attributes represent the last available geo-tags of the considered users, while the labels represent their geo-tags within the next time slot. 

In Figure \ref{fig:percentile} we depict the percentile of the geo-location error. We firstly notice that the percentage of geo-tagged tweets significantly affects the ability of the model to correctly infer the geo-tags. In fact, as expected, the accuracy (measured in terms of distance between the inferred and the true location) is increasing with increasing $p$. This phenomenon is made more evident in Table \ref{tab:mean1km}, where we show the average error and the percentage of tweets that are localized with an error below $1$km with varying $p$. We notice, for example, that the percentage of tweets localized with an error below $1$km drops from $57\%$ to $13\%$ when $p$ goes from $90\%$ to $1\%$. The results summarized in Table \ref{tab:mean1km} suggest that $p=10\%$ represents the threshold to allow an effective localization. In fact, we notice a decrease of about $12$km in the localization error when $p$ goes from $1\%$ to $10\%$, while increasing $p$ from $10\%$ to $30\%$ decreases the error of around $2$ kms only. 

The curves depicted in Fig. \ref{fig:percentile} show similar behaviors. Specifically, the geo-location error slightly increases up to a certain value of percentage of considered tweets (which depends on the geo-tag probability $p$), about which it suddenly increases at a much higher rate. It is therefore possible to divide the tweets into three main categories, namely \textit{highly-predictable}, \textit{average-predictable} and \textit{poorly-predictable} ones. The three categories are depicted in Fig. \ref{fig:percentile_zoom}. The aim of this figure is to underline that our model can accurately localize the largest part of the tweets, which are consequently referred to as highly predictable. Notice that even small values of $p$ allow an efficient geo-localization, i.e., under $1$ km. This confirms the ability of the employed deep learning architecture to efficiently learn common patterns of users' movements. The second category refers to tweets that can be localized with an error that varies between $5$ and $7$ kms if at least $10\%$ of the total tweets are geo-tagged. Finally, the remaining tweets are prone to large localization errors and can be regarded as outliers, whose location cannot be inferred by our model. 

\begin{figure}[!t]
\centering
\includegraphics[width=3in]{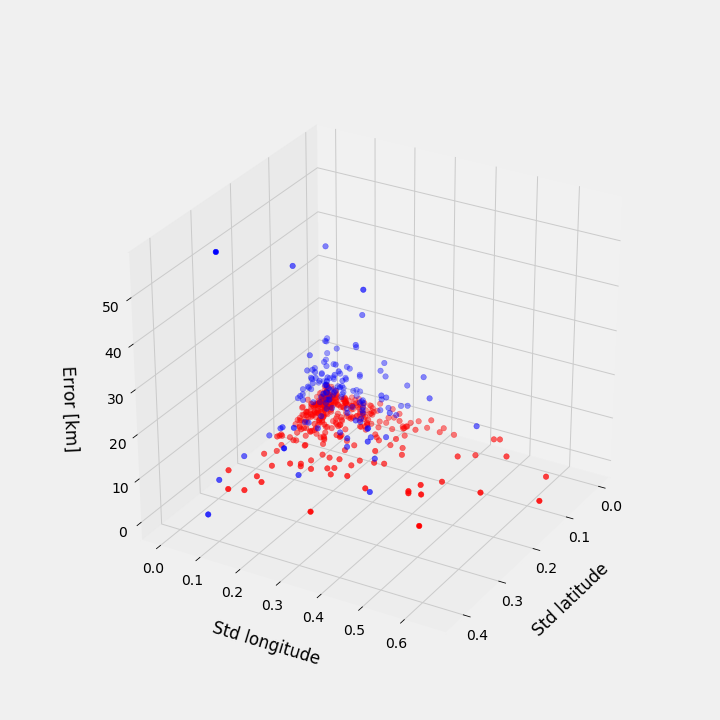} 
\caption{Average localization error (for $p=0.9$) for each user as a function of her mobility dynamics, here represented by the standard deviation of the geo-tagged location coordinates. 
Each point in the figure represents a user in the dataset. Red points represent user with an average localization error below 2.57km, while blue points correspond to users with an error above this value.}
\label{scatter}
\end{figure}
To investigate and better understand the reasons behind these results, we show in Figure \ref{scatter} the average localization error (for $p=0.9$) for each user as a function of her mobility dynamics, here represented by the standard deviation of the geo-tagged location coordinates. 
Each point in the figure represents a user in the dataset. Users with small standard deviations in the latitude-longitude pair indicates \textit{stationary} users, whose mobility is restrained in a small area. On the other side, high standard deviation values characterize \textit{mobile} users.
Red points represent user with an average localization error below 2.57km (according to Table \ref{tab:mean1km}), while blue points correspond to users with an error above this value.
As it can be appreciated from Fig. \ref{scatter}, localization error does not vary with user mobility dynamics.
Though our model is able to infer the majority (73\%) of the users (included those highly mobile) with an accuracy below the average localization error, it fails with some stationary users. Our hypothesis is that these outliers have different mobility patterns compared to the others and, thus, should be modeled separately. We will face this issue in our next work, trying to separate the social network in groups (or communities) according to mobility attributes and locations similarity.

Finally, the overall probability that users provide their tweets with a geo-location (i.e., $p$) is the most important driver of a successful violation of users' privacy. Specifically, the effectiveness of location inference seems to be characterized by a \textit{critical mass}, which is equal to $10\%$ of the total amount of available data in the considered dataset. Notice that the single users do not have this information, as this value results from the aggregation of the single users' probability to geo-tag their tweets. As a future work, we will consider the problem of making the single users able to obtain $p$ in a privacy-preserving fashion (i.e., without asking each user to reveal its probability).

\section{Conclusions}
\label{co}
In this paper, we study the problem of location privacy in Twitter.
In particular, we consider the problem of inferring the current location of a user utilizing only the available geo-tagged tweets from other users in the OSN. We employed a graph-based deep learning architecture to learn a model between the users' known and unknown geo-location during a considered period of time. 

Our experiments validate our approach and further confirms the concern related to data privacy in OSNs.
In fact, one of the main take-away of our work is the presence of a \textit{critical-mass} phenomenon, i.e., if at least $10\%$ of the users provide their tweets with geo-tags, then the privacy of all the remaining users is seriously put at risk. 
Results suggest that with a small percentage of available information (10\% of the geo-tagged tweets) our approach is able to localize almost 50\% of the tweets with an accuracy below 1km. Moreover, we identified a class of highly predictable tweets, whose localization estimate error is really small and does not vary consistently with the percentage of available information in the OSN, further highlighting the weakness of data privacy for some users in the social platform.

Finally, we plan to extend this model in order to improve the inference also for outliers users.
As a future work, we will consider the problem of making users aware of their probability of privacy leakage according to their (and other users) mobility patterns and sharing activities.